\documentclass[a4paper,fleqn,usenatbib,useAMS]{mnras}

\usepackage{graphicx}	
\usepackage{graphics}
\usepackage{amsmath}	
\usepackage{amssymb}	
\usepackage{amsfonts}
\usepackage{multicol}        
\usepackage{multirow}
\usepackage{bm}		
\usepackage{pdflscape}	
\usepackage{natbib}
\usepackage{epstopdf}
\usepackage{enumitem}
\usepackage{soul}

\usepackage[T1]{fontenc}
\usepackage{ae,aecompl}

\title[SZ-induced lensing bias on Planck $\times$ SDSS]{Impact of thermal SZ effect on cross-correlations between \emph{Planck} CMB lensing and SDSS galaxy density fields}

\author[T. Chen, M. Remazeilles]{
Tianyue\,Chen$^{1}$\thanks{E-mail:~\url{tianyue.chen@epfl.ch}},
Mathieu\,Remazeilles$^{2, 3}$\thanks{E-mail:~\url{remazeilles@ifca.unican.es}} 
\\
$^1$Institute of Physics, Laboratory of Astrophysics, Ecole Polytechnique F\'{e}d\'{e}rale de Lausanne (EPFL), Observatoire de Sauverny, \\
1290 Versoix, Switzerland\\
$^2$Instituto de Fisica de Cantabria (CSIC-UC), Avda. los Castros s/n, 39005 Santander, Spain\\
$^3$Jodrell Bank Centre for Astrophysics, Department of Physics and Astronomy, The University of Manchester, Oxford Road, \\
Manchester, M13 9PL, U.K.\\
}

\begin{document}
\label{firstpage}
\pagerange{\pageref{firstpage}--\pageref{lastpage}}
\maketitle


\begin{abstract}
Residual foreground contamination by thermal Sunyaev-Zeldovich (tSZ) effect from galaxy clusters in  cosmic microwave background (CMB) maps propagates into the reconstructed CMB lensing field, and thus biases the intrinsic cross-correlation between CMB lensing and large-scale structure (LSS). 
Through stacking analysis, we show that residual tSZ contamination causes an increment of lensing convergence in the central part of the clusters and a decrement of lensing convergence in the cluster outskirts. 
We quantify the impact of residual tSZ contamination on cross-correlations between the \emph{Planck} 2018 CMB lensing convergence maps and the SDSS-IV galaxy density data through cross-power spectrum computation. 
In contrast with the \emph{Planck} 2018 tSZ-deprojected \textsc{smica} lensing map, our analysis using the tSZ-contaminated  \textsc{smica} lensing map  measures a $\sim2.5\%$  negative bias at multipoles $\ell\lesssim 500$ and transits to a $\sim9\%$ positive bias at $\ell\gtrsim1500$, which validates earlier theoretical predictions of the overall shape of such tSZ-induced spurious cross-correlation. 
The tSZ-induced lensing convergence field  in \emph{Planck} CMB data is detected with more than $1\sigma$ significance at $\ell\lesssim 500$ and more than $14\sigma$ significance at $\ell\gtrsim1500$, yielding an overall $14.8\sigma$ detection.
We also show that masking galaxy clusters in CMB data is not sufficient to eliminate the spurious lensing signal, still detecting a non-negligible bias with $5.5\sigma$ significance on cross-correlations with galaxy density fields. 
Our results emphasize how essential it is to deproject the tSZ effect from CMB maps at the component separation stage and adopt tSZ-free CMB lensing maps for cross-correlations with LSS data.
\end{abstract}

\begin{keywords}
cosmic microwave background -- large-scale structure of Universe -- galaxies: clusters: general -- methods: data analysis -- methods: statistical
\end{keywords}



\section{Introduction}

When travelling along the line-of-sight, the cosmic microwave background (CMB)  photons are deflected by the gravitational potential gradients induced by the large-scale structure (LSS) of the Universe. This is known as the weak gravitational lensing effect \citep[e.g.,][]{Lewis2006}, which distorts the CMB temperature and polarization power spectra, generates spurious CMB $B$-mode polarization, and induces non-Gaussianities through higher-point correlation functions of the lensed CMB field. The CMB lensing effect can be modelled through the integrated lensing potential along the line-of-sight, and is routinely reconstructed from CMB maps by means of quadratic estimators \cite[e.g.,][]{bs87, ce89, Hu2002, cl17}. 

The CMB lensing deflection field indirectly traces  the underlying dark matter distribution in the sky through the imprint of the LSS gravitational potential gradients on the CMB. As a gravitational effect, the lensing distortions of CMB temperature and polarization anisotropies enable to probe dark matter, neutrino masses and dark energy in an unbiased way \cite[e.g.,][]{planck18VIII}. 

Meanwhile, optical galaxy surveys, such as the Baryon Oscillation Spectroscopic Survey (BOSS)/Sloan Digital Sky Survey (SDSS)-III \citep{BOSS2013}, the Dark Energy Survey \citep{des17, des19} and future LSS surveys by the Vera C. Rubin Observatory \citep{LSST2009} and the ESA's \emph{Euclid} satellite \citep{Euclid2011}, directly trace the matter distribution in the Universe, but through a baryonic (or galaxy) bias.

Cross-correlations between CMB lensing maps and optical surveys of galaxy density distribution enable to probe the underlying dark matter while benefiting from mitigated instrumental noise and systematics, which are uncorrelated between two independent data sets. Cross-correlations also offer the advantage of calibrating the galaxy bias of a given tracer \citep{Sherwin2012,Das2013,als+15}, which relate the overdensities of the distribution of galaxies to those of the underlying dark matter distribution. 

For these reasons, many cross-correlation studies between CMB and galaxy surveys are being routinely carried out.
 \cite{smb17} cross-correlated the \emph{Planck} 2015 CMB lensing convergence map with BOSS galaxy catalogues to detect the lensing convergence signal around galaxies out to $100\, h^{-1}{\rm Mpc}$ and measure the galaxy bias. 
 \cite{sms+20} then constrained the amplitude of dark matter fluctuations, $\sigma_8$, and the matter density, $\Omega_m$, in a follow-up paper. 
 \cite{gvh+18} cross-correlated the \emph{Planck} 2015 CMB lensing maps with SDSS galaxy density maps to measure the scale-dependence of the galaxy bias and constrain neutrino masses.
\cite{ogp+19} cross-correlated the combined CMB lensing map from South Pole Telescope (SPT) and \emph{Planck} data with the DES galaxy density map to detect the galaxy-CMB lensing cross-correlation signal and constrain cosmological parameters. \cite{mb19} cross-correlated the combined galaxy clustering catalogue from SDSS, South Galactic Cap u-band Sky Survey (SCUSS), and Wide-field Infrared Survey Explorer (WISE) data with the \emph{Planck} 2018 CMB lensing map, in order to estimate the linear growth of density fluctuations. \cite{syd+21} cross-correlated galaxy groups from the DESI Legacy Imaging Survey with the \emph{Planck} 2018 CMB lensing to constrain the density bias of galaxy groups over various redshifts, mass and richness. 

However, CMB lensing maps may suffer from a spurious lensing signal inherited from the non-Gaussian foreground contamination of the CMB temperature maps by the LSS, such as the residual thermal Sunyaev-Zeldovich (tSZ) emission caused by the hot electron gas in galaxy clusters that scatter off the CMB photons \citep{Sunyaev1972}. Since lensing field estimators are designed to pick up non-Gaussian features in CMB temperature anisotropies  \citep[e.g.][]{Hu2002}, any residual non-Gaussian tSZ contamination in CMB maps at the location of galaxy clusters will induce a \emph{spurious} lensing convergence signal on top of the true CMB lensing convergence field, which may bias the  cross-power spectrum between CMB lensing maps and external LSS tracers.

\cite{van_engelen2014} explored and quantified the bias on the CMB lensing power spectrum induced by tSZ contamination. Through simulations and analytical models, they predicted a negative bias at large angular scales and a positive bias at small angular scales. Large-scale deficit was also observed several times in the cross-power spectrum between galaxy surveys and tSZ-contaminated \emph{Planck} CMB lensing maps \citep[e.g.][]{pah+16,Giannantonio2016}.
With simulated CMB lensing maps, \cite{mh18} demonstrated that the tSZ effect contributes to the most significant bias on CMB lensing$-$galaxy density cross-correlations, and advocated the use of a \emph{tSZ-deprojected} map of CMB temperature anisotropies \citep{Remazeilles2011a} for the reconstruction of the CMB lensing field and its cross-correlation with LSS tracers. By cross-correlating DES galaxy density catalogues with a combined CMB lensing map from \emph{Planck} and SPT data, \cite{boc+19} estimated from a simulation that the tSZ contamination in the CMB lensing map can introduce a large bias on their cross-correlation. 
\cite{ss19} and \cite{ssf20} highlighted large biases induced by various extragalactic foregrounds on CMB lensing$-$galaxy density cross-correlations from simulations of future CMB (Simons Observatory) and LSS (LSST) surveys, and advocated for the use of shear-only and bias-hardening lensing estimators to mitigate these biases in CMB lensing reconstruction.
The aforementioned studies, which partly rely on semi-analytic models and simulations, all predict a typical scale-dependent transitional bias on CMB lensing$-$galaxy cross-correlations caused by the residual tSZ-induced lensing convergence field.

The tSZ contamination of temperature-based CMB lensing maps can thus no longer be ignored, especially given the high sensitivity and resolution of future CMB and LSS surveys. It is essential that we quantify the residual tSZ contamination in existing CMB lensing data sets, understand the induced bias on CMB lensing$-$LSS cross-correlations, and adopt tSZ-free CMB  maps for lensing reconstruction and cross-correlations with LSS whenever possible. 

In a previous paper \citep{crd18}, we studied the impact of the tSZ foreground contamination of the \emph{Planck} 2015 CMB temperature maps by cross-correlation with SDSS, and discussed how tSZ residuals can propagate to the reconstructed CMB lensing map. As a follow-up, in this paper we focus on investigating the impact of tSZ contamination in \emph{Planck} CMB lensing maps through cluster stacking analysis and cross-correlation with SDSS galaxy densities. We use the latest \emph{Planck} 2018 lensing data set \citep{planck18VIII}, where both tSZ-contaminated and tSZ-free lensing products are available for comparison. The difference between these two \emph{Planck} lensing products allows us for the first time to get the spurious lensing convergence field induced by residual tSZ foreground contamination directly from the data, and to study its impact on cross-correlations with LSS data from SDSS.

This paper is organised as follows. Section~\ref{theory} outlines the theoretical framework for this work. Section~\ref{datasec} describes the data sets used in our analysis. Section~\ref{sec:stack} shows the results of stacking analysis in map space.  Section~\ref{cross_spec} presents the results from the cross-power spectrum analysis. Section~\ref{dissec} summarises the results and draws the conclusions.


\section{Theoretical framework}\label{theory}

Gravitational lensing by LSS distorts the primary CMB temperature anisotropies through the following remapping \citep[e.g.,][]{Lewis2006}:
\begin{align}\label{eq:cmblensing}
T_{\rm CMB}(\mathbf{\hat{n}}) &= T_{\rm CMB}^{\rm unl}(\mathbf{\hat{n}}+\nabla\Phi(\mathbf{\hat{n}}))\cr
&\simeq T_{\rm CMB}^{\rm unl}(\mathbf{\hat{n}})+\nabla\Phi(\mathbf{\hat{n}})\cdot\nabla T_{\rm CMB}^{\rm unl}(\mathbf{\hat{n}})\,,
\end{align}
where $T_{\rm CMB}(\mathbf{\hat{n}})$ is the observed CMB temperature anisotropy in the direction $\mathbf{\hat{n}}$, $T_{\rm CMB}^{\rm unl}(\mathbf{\hat{n}})$ is the unlensed primary CMB temperature anisotropy, and $\Phi(\mathbf{\hat{n}})$ is the lensing potential whose gradients induce spatial distortions to the primary CMB temperature anisotropies. For a flat Universe, the lensing potential is  defined by
 \begin{equation}\label{eq:potential}
   \Phi(\mathbf{\hat{n}}) = -2\int_0^{\chi_*}d\chi\frac{\chi_*-\chi}{\chi_*\chi}\Psi(\chi\mathbf{\hat{n}};\eta_0-\chi)\,,
\end{equation}
 where $\chi$ is the comoving distance, $\chi_*$ is the distance to the CMB last-scattering surface, and $\Psi(\chi\mathbf{\hat{n}};\eta_0-\chi)$ is the gravitational potential of the LSS at comoving distance $\chi$ along the direction $\hat{n}$, and $\eta_0$ is the conformal time today. Aside from the lensing potential Eq.~\eqref{eq:potential}, additional observables can be used to describe the CMB lensing effect: the deflection vector $\mathbf{d}(\mathbf{\hat{n}})=\nabla\Phi(\mathbf{\hat{n}})$ as the gradient of the lensing potential, and the convergence field  $\kappa(\mathbf{\hat{n}})={1\over 2} \nabla^2\Phi(\mathbf{\hat{n}})$ as the divergence of the deflection vector, i.e. the Laplacian of the lensing potential.
 
 Taking the spherical harmonic transform of Eq.~\eqref{eq:cmblensing}, the lensing potential can be estimated from the observed CMB temperature anisotropies by means of quadratic estimators which perform a weighted convolution in harmonic space of the CMB map with itself \citep[e.g.,][]{oh03}:
\begin{equation}\label{eq:qe}
 \widehat{\Phi}(\mathbf{L}) \propto  \int \frac{d^2\mathbf{l}}{(2\pi)^2}\,F(\mathbf{l}, \mathbf{L-l})\,T_{\rm CMB}(\mathbf{l})\,T_{\rm CMB}(\mathbf{L-l})\,,
\end{equation}
where $T_{\rm CMB}(\mathbf{l})$ is the spherical harmonic coefficient of the observed CMB map at multipole $\mathbf{l} = (\ell, m)$, and
\begin{equation}
  F(\mathbf{l}, \mathbf{L-l}) = \frac{\mathbf{L} \cdot \mathbf{l}C^{\rm unl}_{\ell}+\mathbf{L}\cdot\mathbf{(L-l)}C^{\rm unl}_{|\mathbf{L-l}|}}{(C^{\rm unl}_{\ell}+N_{\ell})\left(C^{\rm unl}_{|\mathbf{L-l}|}+N_{|\mathbf{L-l}|}\right)}
\end{equation}
is the minimum-variance filtering function that downweights the noise,  with $C_{\ell}^{\rm unl}$ and $N_{\ell}$ being the assumed unlensed CMB and noise power spectra.  As the Laplacian of the lensing potential, the lensing convergence field is trivially derived in harmonic space from the reconstructed lensing potential Eq.~\eqref{eq:qe} as
\begin{equation}
  \widehat{\kappa}_{\ell m} = \frac{\ell(\ell+1)}{2}\widehat{\Phi}_{\ell m}\,.
  \label{eq:kappa}
\end{equation}

As evident from Eq.~\eqref{eq:qe}, quadratic estimators rely on the breaking of the statistical isotropy of the primary CMB anisotropies by the lensing distortion, i.e. $\langle T_{\rm CMB}(\mathbf{l})\,T_{\rm CMB}(\mathbf{L-l})\rangle \neq 0$ for $\mathbf{L}\neq 0$. Hence, the estimated lensing fields  $\widehat{\Phi}$ and $\widehat{\kappa}$ may inherit from residual contamination by non-Gaussian foreground signals which are left in CMB temperature maps after component separation, such as the residual tSZ emission in \emph{Planck} CMB maps in the direction of galaxy clusters \citep[see e.g.][]{crd18}. 

Component separation methods extract a cleaned CMB map from multi-frequency observations through some weighted combination of the frequency maps aimed at mitigating astrophysical foreground emissions \citep[see e.g.][for a review]{dc07}. Concerning the \emph{Planck} \textsc{smica} CMB temperature map \citep{planck18IV}, from which the \emph{Planck} CMB lensing maps have been derived \citep{planck18VIII}, the combination of the \emph{Planck} frequency maps is performed in harmonic space, i.e.
\begin{equation}
   \widehat{T}_{{\rm CMB},\, \ell m}= \sum_{\nu}w_{\ell, \nu}\,x_{\ell m, \nu}\,,
\end{equation}
where the sum runs over the \emph{Planck} frequency channels $\nu$, $w_{\ell, \nu}$ are multipole-dependent weights, and $x_{\ell m, \nu}$ are the spherical harmonic coefficients of the \emph{Planck} frequency maps which include emission from the CMB but also foreground emissions such as the tSZ effect:
\begin{equation}\label{eq:obs}
\,x_{\ell m, \nu} = a_\nu\, T_{{\rm CMB},\, \ell m} + b_\nu\, y_{\ell m} + \cdots
\end{equation}
In Eq.~\eqref{eq:obs} $a_\nu$ is the spectral energy distribution (SED), or spectral response, of the CMB anisotropies across the frequencies (i.e. temperature derivative of blackbody), while $b_\nu$ is the SED of the tSZ emission and $y_{\ell m}$ the spherical harmonic coefficients of the tSZ Compton-$y$ parameter. 
The multipole-dependent weights $w_{\ell, \nu}$ are constrained to give unit gain to the CMB signal, i.e. $\sum_{\nu}w_{\ell, \nu}\,a_\nu=1$, while guaranteeing minimum variance from the other signals. However, minimum-variance weights are generally not orthogonal to the tSZ SED vector, i.e. $\sum_{\nu}w_{\ell, \nu}\,b_\nu \neq 0$, so that CMB temperature map estimates receive some mitigated but non-zero residual contribution from the tSZ effect:
\begin{equation}
   \widehat{T}_{{\rm CMB},\, \ell m}= T_{{\rm CMB},\, \ell m} + \left(\sum_{\nu}w_{\ell, \nu}\,b_\nu\right)\,y_{\ell m}+\cdots \,,
\end{equation}
As a statistically anisotropic and non-Gaussian field, the residual emission from tSZ effect propagates into the reconstructed lensing field through the lensing estimator Eq.~\eqref{eq:qe}, thus effectively biasing the intrinsic CMB lensing field by adding an extra spurious, tSZ-induced, lensing field:
\begin{equation}\label{eq:sz_lensing}
   \widehat{\kappa}_{\rm SZ}(\mathbf{L}) \propto \Delta\widehat{\Phi}_{\rm SZ}(\mathbf{L}) \propto  \int \frac{d^2\mathbf{l}}{(2\pi)^2}\,\widetilde{F}(\mathbf{l}, \mathbf{L-l})\,y(\mathbf{l})\,y(\mathbf{L-l})\,,
 \end{equation}
where $y(\mathbf{l})$ is the spherical harmonic coefficient of the Compton-$y$ parameter at multipole $\mathbf{l} = (\ell, m)$, and 
\begin{equation}\label{eq:sz_kernel}
    \widetilde{F}(\mathbf{l}, \mathbf{L-l}) = \left(\sum_{\nu} w_{\mathbf{l}, \nu}\,b_{\nu}\right)F(\mathbf{l}, \mathbf{L-l})\left(\sum_{\nu} w_{\mathbf{L-l}, \nu}\,b_{\nu}\right)\,.
\end{equation}

The tSZ effect and the optical galaxy surveys trace the same underlying non-Gaussian matter distribution. Therefore, cross-correlations $\langle \kappa\,, \delta_g\rangle$ between CMB lensing convergence maps $\kappa$ and galaxy density fields $ \delta_g$ must be biased due to the tSZ-induced spurious lensing convergence field $\kappa_{\rm SZ}$ (Eq.~\ref{eq:sz_lensing}) which correlates with the galaxy density field because of non-zero three-point functions of the underlying non-Gaussian matter distribution:
\begin{equation}\label{eq:bias}
\langle \kappa_{\rm SZ}\,, \delta_g\rangle \propto  \int \frac{d^2\mathbf{l}}{(2\pi)^2}\,\widetilde{F}(\mathbf{l}, \mathbf{L-l})\,\langle y(\mathbf{l})\,y(\mathbf{L-l})\,\delta_g(\mathbf{L})\rangle\,.
 \end{equation}
 Similar formulas of spurious three-point correlations have been derived for other extragalactic foregrounds \citep[e.g.][]{Ferraro2018}.
 As shown by several authors on semi-analytical models and simulations, the tSZ-induced bias Eq.~\eqref{eq:bias} on CMB lensing-LSS cross-correlations has a typical scale-dependent sign-changing shape across the multipoles that can be quite significant \citep[e.g.][]{van_engelen2014, mh18,boc+19}.

While masking galaxy clusters in CMB temperature maps prior to lensing reconstruction might help mitigating such spurious tSZ-induced lensing signal, it can introduce another bias to the lensing field estimate because of the effective spatial correlation between the extragalactic-source mask and the expected peaks of the CMB lensing convergence field \citep{Fabbian2021}. 

A robust alternative to get rid of the spurious tSZ-induced lensing field Eq.~\eqref{eq:sz_lensing} in the CMB lensing estimation, and thus eliminate biases Eq.~\eqref{eq:bias} in CMB lensing-LSS cross-correlations, is to fully \emph{deproject} the tSZ effect from the reconstructed CMB temperature map at the component separation stage \citep{Remazeilles2011a}. This is achieved by a constrained internal linear combination (hereafter, Constrained ILC) which relaxes the minimum-variance condition to allow the component separation weights to be orthogonal to the tSZ SED, i.e. 
\begin{equation}\label{eq:cilc}
\sum_{\nu}w_{\ell, \nu}\,b_\nu = 0\,,
 \end{equation}
 in addition to satisfying the constraint of CMB signal conservation, i.e. $\sum_{\nu}w_{\ell, \nu}\,a_\nu = 1$. Using the Constrained ILC weighting Eq.~\eqref{eq:cilc}, a tSZ-free version of the \textsc{smica} CMB temperature map has been released by the \emph{Planck} collaboration \citep{planck18IV} for the final 2018 data release, and used as a tSZ-free alternative for CMB lensing field reconstruction \citep{planck18VIII}. The Constrained ILC method is also now routinely used by other CMB collaborations to deliver tSZ-free CMB temperature maps \citep[e.g.][]{Madhavacheril2020} and thereby guarantee unbiased CMB lensing maps \citep[e.g.][]{Darwish2021}.  The noise variance increase on CMB lensing estimation which results from the additional tSZ deprojection constraint Eq.~\eqref{eq:cilc} can in principle be mitigated by using the tSZ-free CMB map in only one of the two legs of the lensing quadratic estimator \citep{mh18}.


\section{Data sets}\label{datasec}

In this section, we describe the set of CMB lensing products (Section\;\ref{seclensing}) and LSS data (Sections~\ref{bossdata} and~\ref{szcat}) used in our cross-correlation analysis.

\subsection{\textit{Planck} lensing maps} \label{seclensing}

 We use three variations of CMB lensing products from the \emph{Planck} 2018 data release\footnote{\url{https://wiki.cosmos.esa.int/planck-legacy-archive/index.php/Lensing}} \citep{planck18VIII}:
 \begin{enumerate}[label=(\roman*)]
 \item The  tSZ-contaminated \emph{Planck} \textsc{smica} CMB lensing convergence map, $\kappa_{\rm SMICA}$.
 \item  The tSZ-masked \emph{Planck} CMB lensing convergence map, $\kappa_{\rm Cluster-masked}$, derived from the \textsc{smica} CMB temperature map in which tSZ clusters have been masked out.
 \item A tSZ-free version of the \emph{Planck} CMB lensing convergence map, $\kappa_{\rm SMICA-noSZ}$, derived from the so-called \textsc{smica-nosz} CMB temperature map in which the tSZ effect has been deprojected according to Eq.~\eqref{eq:cilc}.
 \end{enumerate}
 We use the temperature-only lensing estimate from each product, since polarisation is irrelevant to our analysis.  
 
 The \emph{Planck} 2018  release gives the lensing products as the convergence field $\kappa_{\ell m}$ defined by Eq.~\eqref{eq:kappa} with a maximum multipole of $\ell_{\rm max} = 4096$.
For all three variations, we map the convergence $\kappa_{\ell m}$ on the sphere through a \textsc{healpix} pixelization scheme \citep{ghb+05} with $N_{\rm side}=2048$ (pixel size of $\sim1.7$\,arcmin).  The $\kappa$ map is convolved with a Gaussian beam window of $5$\,arcmin FWHM in order to be consistent with the angular resolution of the \emph{Planck} satellite.

 Taking the difference between the tSZ-contaminated \emph{Planck} lensing $\kappa_{\rm SMICA}$ map and the tSZ-deprojected \emph{Planck} lensing $\kappa_{\rm SMICA-noSZ}$ map allows us to derive the tSZ-induced spurious lensing convergence field (Eqs.~\ref{eq:sz_lensing}-\ref{eq:sz_kernel}) directly from the \emph{Planck} data:
\begin{equation}\label{eq:kappasz}
\kappa_{\rm SZ} \equiv \kappa_{\rm SMICA} - \kappa_{\rm SMICA-noSZ}\,.
\end{equation} 
Cross-correlating this residual tSZ lensing field with SDSS galaxy density data will allow us to quantify the tSZ-induced lensing bias on CMB lensing-galaxy cross-correlations. We note that the map difference in Eq.~\eqref{eq:kappasz} may exhibit extra contributions from other extragalactic foregrounds such as the cosmic infrared background (CIB), since the constraint of tSZ deprojection in one of the two maps can impact the mitigation of the variance of other foregrounds by requisitioning one degree of freedom during component separation \citep{Remazeilles2011a}, and thus may leave slightly more CIB contamination in the tSZ-deprojected map. However, such potential CIB penalty resulting from tSZ deprojection mostly affects CMB experiments with a few number of frequency channels and limited high-frequency coverage from the ground \citep{Abylkairov2021,ssf+21}. This is not the case for \emph{Planck}, for which the map difference in Eq.~\eqref{eq:kappasz} is largely dominated by the tSZ effect as we already showed in \cite{crd18}.

In addition to Eq.~\eqref{eq:kappasz}, we also consider the difference between the tSZ-masked \emph{Planck} lensing $\kappa_{\rm Cluster-masked}$ map and the tSZ-deprojected \emph{Planck} lensing $\kappa_{\rm SMICA-noSZ}$ map, in order to assess whether masking tSZ clusters in the CMB map prior to lensing reconstruction still leaves a residual tSZ-induced lensing bias on CMB lensing-galaxy cross-correlations.

For the \emph{Planck} 2018 lensing products,  a combination of masks has been used: the \textsc{smica}-based confidence mask, a $70\%$ Galactic mask and \emph{Planck} point-source masks at 143\,GHz and 217\,GHz. For the tSZ-masked \emph{Planck} lensing $\kappa_{\rm Cluster-masked}$ map, an additional mask is applied to remove the resolved tSZ clusters detected at ${\rm SNR} > 5$ in the \emph{Planck} 2015 SZ catalogue \citep{planckXXVII}, leaving a total fraction of observed sky of $f_{\rm sky}=0.67$. 


\subsection{SDSS catalogue}\label{bossdata}

We use the same LSS data set in our analysis as in \cite{crd18},  produced from the main photometric galaxy (MphG) catalogue given by SDSS-IV  survey Data Release 13 \citep[DR13;][]{aaa+17}. The MphG catalogue provides galaxy magnitudes within five optical filter bands $u$, $g$, $r$, $i$, $z$. We use the $r$ band magnitude as our reference for sample selection due to its better sensitivity and  calibration accuracy. Faint sources with magnitude below the completeness level ($r>22.2$) are excluded from our sample since they can introduce uncorrelated background noise during cross-correlation by smearing the true signal.  Also discarded are the top $1\%$ brightest sources with magnitude $r<17$,  in order to avoid this small amount of bright sources dominating the statistical results. With our selection criteria, a total of $\sim$133 million galaxies ($\sim64\%$ of the full MphG catalogue) are included in our sample,  covering a sky area of $14555$ square degrees.

According to the source coordinate provided by the MphG catalogue, each selected source is assigned into a single \textsc{healpix} pixel with $N_{\rm side} = 2048$, since the typical angular size of SDSS galaxies is much smaller than the $\sim1.7$\,arcmin pixel size \citep{slb+02}. The SDSS galaxy density contrast map is constructed as
\begin{equation}\label{eq:contrast}
\delta_g = \frac{N-\overline{N}}{\overline{N}}\,,
\end{equation}
where  $N$ is the number of sources in each pixel and $\bar{N}$ is the average number of sources per pixel. The density contrast map is then convolved with a 5\,arcmin beam to be consistent with the angular resolution of the \emph{Planck} CMB lensing maps. 

\subsection{\emph{Planck} tSZ maps}\label{szcat}

In addition to SDSS galaxy density data, we consider \emph{Planck} tSZ maps as another independent tracer of the LSS. The \emph{Planck} tSZ maps trace galaxy clusters which contribute to part of the lensing potential, and thereby \emph{Planck} tSZ maps must be correlated with \emph{Planck} CMB lensing maps. Therefore, it is also important to quantify the bias on CMB lensing-tSZ cross-correlations caused by the spurious tSZ foreground-induced lensing field. Since the galaxy clusters of the \emph{Planck} tSZ maps host some of the galaxies probed by SDSS, the bias on CMB lensing-tSZ cross-correlations is expected to have the same overall shape across multipoles than that of the bias on CMB lensing-galaxy cross-correlations.

We make use of the \emph{Planck} \textsc{nilc} tSZ Compton-$y$ map \citep{planck2015sz} and another tSZ Compton-$y$ map of confirmed clusters simulated from the \emph{Planck} 2015 SZ catalogue  \citep{planckXXVII}. The former, which we note $y$, traces the hot gas of electrons in the entire sky, i.e. in all galaxy clusters and diffuse filaments, but is quite noisy, while the latter, which we note $y^{\rm sim}$, traces only the compact galaxy clusters of the \emph{Planck} SZ catalogue detected at signal-to-noise ratio $\rm{SNR} > 4.5$.

The \emph{Planck} SZ catalogue provides the integrated tSZ flux $Y_{5R500}$ within a radius of $5\times R_{500}$ and the angular size $\theta_s$ of each cluster. Since the majority of these clusters have their angular sizes smaller than the \emph{Planck} beam size of $5$\,arcmin, one can assume that each cluster is a flat disc with radius $\theta_s$ and uniform brightness. Under this assumption, one can derive the Compton parameter $\hat{y}^{\rm sim}$ from $Y_{5R500}$ and $\theta_s$ through
\begin{equation}
  Y_{5R500} \simeq \hat{y}^{\rm sim}\times \pi\theta_s^2\,.
\end{equation}
The clusters are projected onto a \textsc{healpix} map with ${N_{\rm side}  = 2048}$ according to their sky locations given by the \emph{Planck} SZ catalogue.  The catalogue SZ map is then smoothed to 5\,arcmin to be consistent with the angular resolution of the \emph{Planck} CMB lensing maps. 

The \emph{Planck} \textsc{nilc} $y$-map has an angular resolution of $10$ arcmin. Therefore, for cross-correlations involving the \emph{Planck} \textsc{nilc} $y$-map we smooth the \emph{Planck} CMB lensing convergence maps to the same $10$ arcmin resolution.


\subsection{Simulations}\label{simsec}

In order to interpret the observed correlation excess and/or deficit between CMB lensing and LSS surveys caused by tSZ residuals in CMB maps and obtain sample uncertainties, we also simulate a set of  Monte Carlo (MC) CMB lensing maps. 

We first compute a theoretical CMB power spectrum from \textsc{camb} \citep{aa11, hlh+12} based on the \emph{Planck} 2018 $\Lambda$CDM model \citep{planckVI18}, and generate 1000 Gaussian realisations of pure CMB temperature maps from  this theoretical spectrum using the \textsc{synfast} facility in \textsc{healpix} \citep{ghb+05}. In addition, we generate 1000 MC CMB temperature maps with artificial tSZ contamination by adding to the pure CMB maps the tSZ catalogue map described in Sect.\,\ref{szcat} which we scale to $143$\,GHz using the tSZ spectral energy distribution \citep[see][]{crd18}.

We then lens both the pure and tSZ-contaminated MC CMB maps with a lensing potential field using the \textsc{LensIt} code\footnote{\url{https://github.com/carronj/LensIt}} \citep{cl17}. The code makes use of \textsc{camb} to  compute a lensing potential power spectrum $C_{\ell}^{\phi\phi}$, and uses the \textsc{synalm} facility in \textsc{healpix}  to generate a set of spherical harmonic coefficients $\phi_{\ell m}$, corresponding to the lensing potential. The $\phi_{\ell m}$ is then transformed into  a spin-1 deflection field $d_{\ell m}  = \sqrt{\ell(\ell+1)}\phi_{\ell m}$ to lens  the unlensed MC CMB temperature maps  in our case.

Finally we reconstruct the lensing convergence field from the lensed MC CMB maps using the \emph{Planck} 2018 lensing pipeline\footnote{\url{https://github.com/carronj/plancklens}} \citep{planck18VIII}. For each lensed CMB map,  the pipeline first computes the lensing gradient estimator, which is then normalised by the expected normalization in the fiducial cosmology to calculate the lensing potential $\phi_{\ell m}$. The lensing convergence field $\kappa_{\ell m}$ is calculated from $\phi_{\ell m}$ through Eq.~\eqref{eq:kappa} and transformed to a sky map using the \textsc{alm2map} facility in \textsc{healpix}. Finally, the convergence maps are smoothed to 5\,arcmin to be consistent with the \emph{Planck} 2018 lensing maps. In summary, we have two sets of MC simulated lensing convergence maps, generated from 1000 pure CMB and artificially tSZ-contaminated CMB  temperature maps respectively. 

\begin{figure*}
  \includegraphics[width=1\hsize]{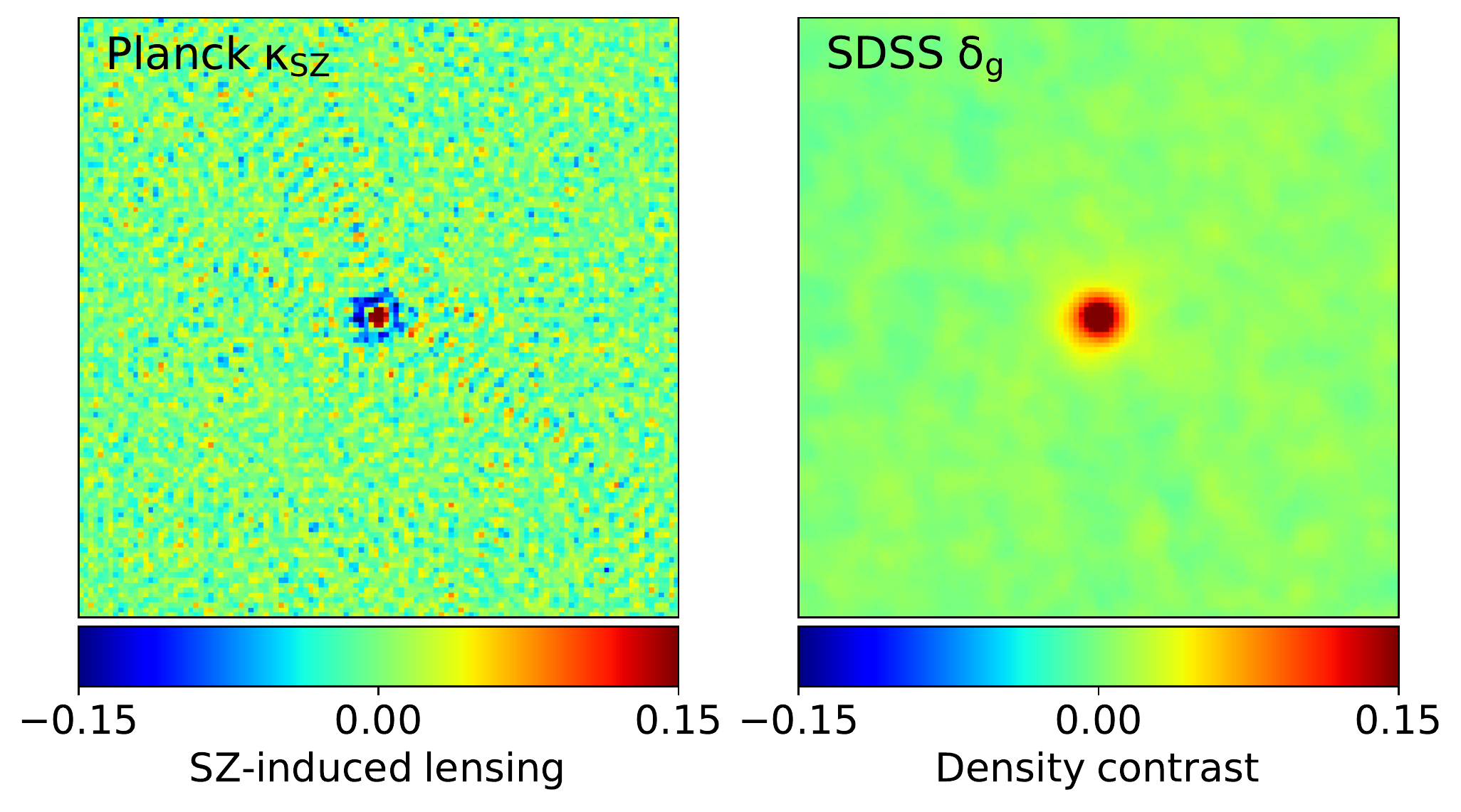}
  \includegraphics[width=1\hsize]{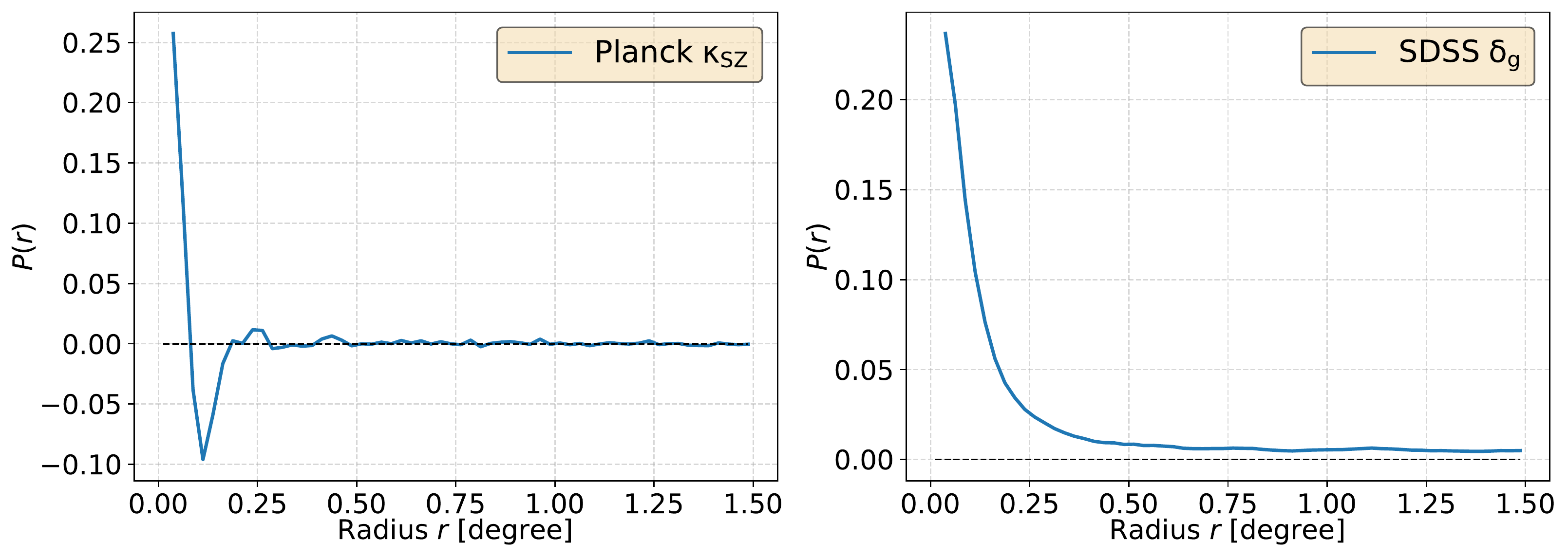}
\caption{\emph{Upper panels}: The stacked maps of $\kappa_{\rm SZ}$ (\emph{left}) and $\delta_g$ (\emph{right}) at the centre locations of galaxy clusters from the \emph{Planck} 2015 SZ catalogue  \protect\citep{planckXXVII}. Each map is $3^{\circ}\times3^{\circ}$ with an angular resolution of 5\,arcmin. \emph{Lower panels}: The radial profile of the stacked maps as  a function of radius from the centre. In each row, \emph{left}: The residual tSZ-induced lensing convergence; \emph{right:} The SDSS galaxy density contrast. Residual tSZ contamination in the CMB temperature map causes an increment of lensing convergence in the central part of the clusters and a decrement of lensing convergence in the outskirts of the clusters.} 
\label{fig:stack}
\end{figure*}
 
\section{Stacking analysis}\label{sec:stack}
In this section, we stack the tSZ-induced lensing convergence field $\kappa_{\rm SZ}$ (Eqs.~\ref{eq:sz_lensing}, ~\ref{eq:kappasz}) and the SDSS density contrast map $\delta_g$ (Eq.~\ref{eq:contrast}) at the locations of SZ galaxy clusters according to the \emph{Planck} 2015 SZ catalogue  \citep{planckXXVII}. We visually inspect the stacked maps and compute the stacking profiles for a first sight of correlation.

\subsection{Map stacking}
In order to demonstrate the correlation between the tSZ-induced lensing field and the SDSS galaxies, we stack the residual tSZ lensing convergence map, $\kappa_{\rm SZ}$, defined in  Eq.~\eqref{eq:kappasz} at the locations of the galaxy clusters of the \emph{Planck} 2015 SZ catalogue  \citep{planckXXVII}. We extract from the $\kappa_{\rm SZ}$ map $3^{\circ}\times3^{\circ}$ patches of the sky with $120\times120$ pixels of 1.5\,arcmin pixel size centred around each cluster of the catalogue. We then average all the patches to get the stacked residual tSZ-induced lensing field shown in the upper left panel of Fig.~\ref{fig:stack}.  Clear positive temperature fluctuations are detected at the centre of the SZ cluster locations shown as the central red spot. It is surrounded by negative temperature fluctuations shown as the outer blue rings. The stacked $\kappa_{\rm SZ}$ map demonstrates that residual tSZ emission in the CMB map induces an \emph{increment }of lensing convergence at small scales in the centre of the clusters,  and a \emph{decrement} of lensing convergence at larger scales in the outskirts of the clusters.   

Similarly, we stack the SDSS density contrast map at the locations of SZ clusters as shown in the upper right panel of Fig.\,\ref{fig:stack}.  A positive overdensity is detected (red spot) at cluster locations.  By comparing the two upper panels in Fig.\,\ref{fig:stack}, our stacking results visually demonstrate that the tSZ-induced lensing convergence field is correlated with LSS. A positive correlation is expected at small scales centred at the clusters, and an anti-correlation is expected at larger scales  due to the negative lensing field at the boundary of clusters. This peculiar correlation signature is shown for the first time on maps.

\subsection{Radial profile}
To further confirm the scale-dependent correlation seen in the stacked maps, we compute the radial profile of the stacked maps defined as
\begin{equation}\label{equ:profile}
P(r) = \frac{\sum_{r-\Delta r/2}^{r+\Delta r/2}\sum_{\theta = 0}^{\theta = 2\pi}I(\theta', r')}{N_{\rm pix}}\,,
\end{equation}
where $I(\theta', r')$ is the amplitude of the stacked map at each pixel within annuli of inner radius $r-\Delta r/2$ and outer radius $r+\Delta r/2$ from the centre. $N_{\rm pix}$ is the total number of pixels within each annulus.  $P(r)$ quantifies the average amplitude of the stacked maps as a function of the angular radius from the centre. In our case, we have the radius ranging between $0^{\circ}$ and $1.5^{\circ}$ corresponding to the $3^{\circ}\times3^{\circ}$ stacked maps. We choose the bin size to be $\Delta r = 1.5$\,arcmin, corresponding to a single pixel resolution. Therefore, we calculated $P(r)$ at 60 equally spaced points between the radius of $0^{\circ}$ and $1.5^{\circ}$. 

The calculated radial profiles are shown in the lower panels of Fig.\,\ref{fig:stack} for the tSZ-induced lensing convergence field (\emph{left}) and the SDSS density contrast  map (\emph{right}). The radial profile of the stacked $\kappa_{\rm SZ}$ map (\emph{lower left}) shows an average of positive amplitude within a $\sim0.1^{\circ}$ distance from the SZ galaxy centres. Further than $\sim0.1^{\circ}$, the average amplitude drops to negative and flattens out beyond the cluster scale at $r\gtrsim0.5^{\circ}$. The profile reflects the positive red spot surrounded by the negative blue ring in the stacked map  as seen in the upper left panel. In comparison, the radial profile of the  stacked SDSS map (\emph{lower right}) is always positive  as seen from the red spot in the stacked SDSS map in the upper right panel.

The radial profiles further highlight the peculiar correlation between tSZ-induced lensing convergence field and LSS data in the direction of galaxy clusters.  In particular, the characteristic transition of sign is addressed by comparing the two radial profiles in the lower panels of Fig.\,\ref{fig:stack}. This new result obtained from \emph{Planck} and SDSS maps stacking corroborates previous literature using model simulations and power spectrum analyses \citep{mh18, van_engelen2014}, which predicted a scale-dependent transitional shape of the tSZ-induced bias in the CMB lensing-galaxy cross-power spectrum. 

\section{Cross-power spectrum analysis} \label{cross_spec}

In this section, we compute cross-correlations between LSS data and \emph{Planck} CMB lensing maps in order to detect the bias caused by the spurious lensing field induced by tSZ residuals in CMB data. The same set of masks as in \cite{crd18} is applied to our maps  before computing the cross-spectrum: a Galactic mask excluding pixels at Galactic latitude $|b|<30^{\circ}$ to mitigate Galactic foreground contamination in \emph{Planck} CMB lensing maps and an SDSS mask to exclude the unobserved sky region from SDSS. 
The combined mask is apodized by  a $80'$ beam to  smooth out the sharp transition at the boundary and thereby avoid artefacts arising from spherical harmonic transforms when computing angular power spectra. Afterwards, pixels of the apodized mask with a value below the threshold of $0.5$ are set to zero. This mask leaves a total fraction of sky coverage of $f_{\rm sky} = 27\%$.

We use the estimator \textsc{polspice} \citep{spc01,ccp+04,  cc05} for calculating angular cross-power spectra and correcting for the effects of mask, pixel window function and beam window function. We compute the cross-power spectra up to a maximum multipole value of $\ell_{\rm max} = 4096$ according to the \emph{Planck} pixel resolution.  Due to significant impact from mask and apodization,  the first bin of 100 multipoles ($\ell = 2-100$) are excluded from our analysis. The cross-power spectra are divided into 5 other bins of multipoles: $100<\ell<500$, $500<\ell<1500$, $1500<\ell<2500$, $2500<\ell<3500$, and $3500<\ell<4000$.

\begin{figure}
\includegraphics[width=1.0\hsize]{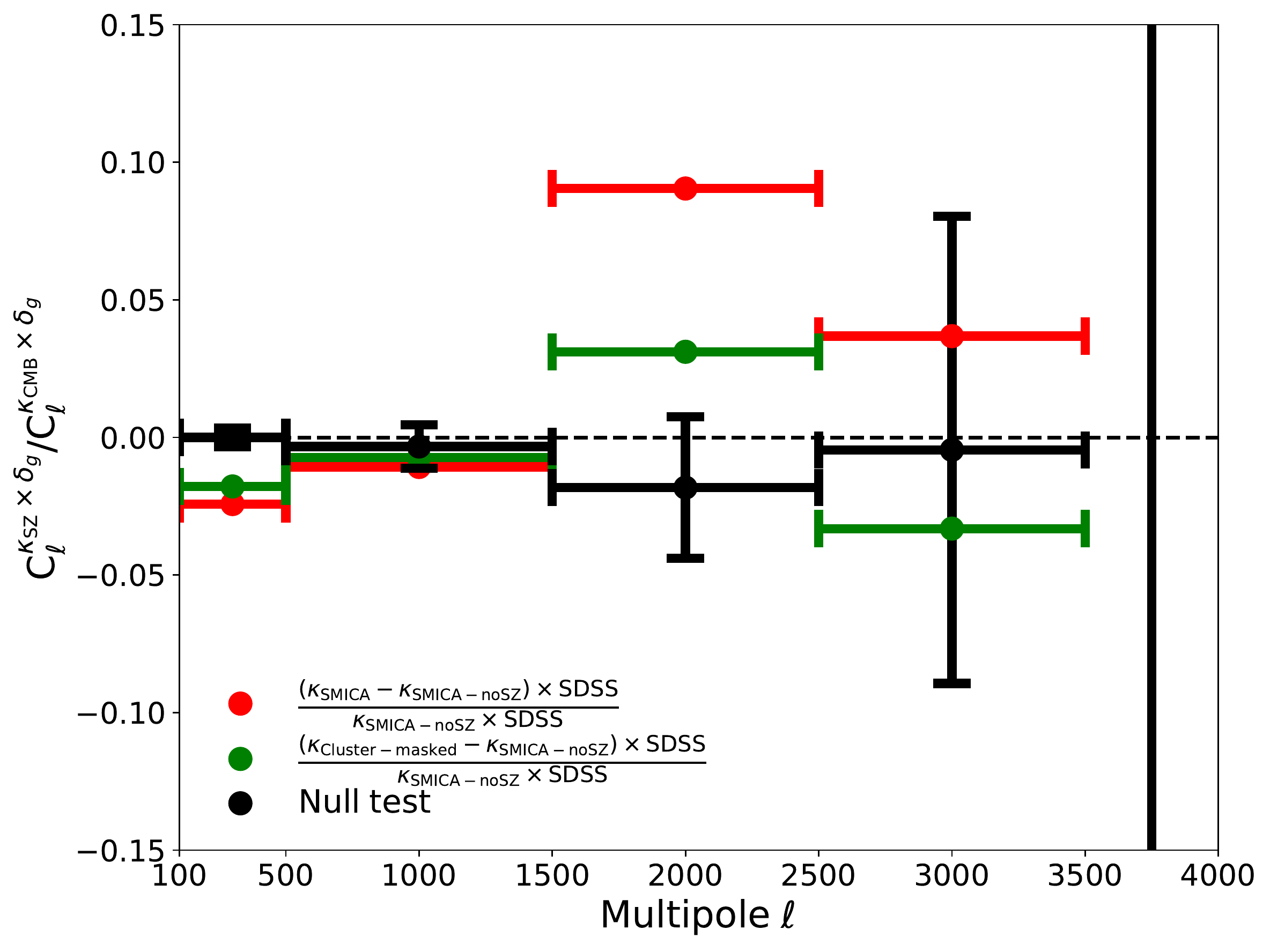}
\caption{The cross-power spectrum $C_{\ell}^{\kappa_{\rm SZ}\times \delta_g}$ relative to  $C_{\ell}^{\kappa_{\rm CMB}\times \delta_g}$, replicating Fig.~4 in \protect\cite{mh18} using the \emph{Planck} 2018 lensing dataset  \citep{planck18VIII}. The $\kappa_{\rm SZ}$ map in the \emph{red} data set is from the difference between the tSZ-contaminated \textsc{SMICA} and tSZ-deprojected \textsc{SMICA} lensing  maps. The $\kappa_{\rm SZ}$ map in the \emph{green} data set is from the difference between the cluster-masked  \textsc{SMICA} and SZ-deprojected \textsc{SMICA} lensing  maps. In both cases, the $\kappa_{\rm SZ}$ map is cross-correlated with the SDSS galaxy density map. The horizontal bars indicate the multipole bin width. The black error bars give the average and $1\sigma$ sample variance computed from 1000 MC simulations with randomised SDSS galaxy locations.} 
\label{fig:reMH18}
\end{figure}

\subsection{SZ-induced lensing bias on the $\kappa_{\rm CMB} \times {\rm SDSS}$  cross-power spectrum}
\label{sec-kxsdss}

\begin{table}
\centering
\begin{tabular}{l|cc}
\hline
\hline
\multirow{2}{*}{Multipole Range} & \multicolumn{2}{c}{S/N}  \\
& tSZ-contaminated & Cluster-masked \\
\hline
{[100, 500]} & 7.4 & 5.4\\
{[500, 1500]} & 1.0 & 0.7\\
{[1500, 2500]} & 12.0 & 4.1\\
{[2500, 3500]} & 0.5 & 3.7\\
{[3500, 4000]} & 3.5 & 2.7\\
\hline
[100, 4000] & 14.6 & 7.4\\
\hline
\hline
\end{tabular}
\caption{Detection significance of the correlation bias on the \emph{Planck} CMB lensing-SDSS galaxy cross-power spectrum due to \emph{Planck} tSZ-induced lensing contamination over different multipole ranges. The S/N is based on Fig.~\ref{fig:reMH18}. The S/N in the middle column is computed using the tSZ-contaminated lensing map, while the last column is from the cluster-masked lensing map.}
\label{tab:MH18SN}
\end{table}

We first assess the amplitude of the bias on the CMB lensing-galaxy density cross-correlation that is caused by the residual tSZ-induced lensing field.  To this end we replicate the figure 4 of \cite{mh18} using \emph{Planck} lensing maps and SDSS data, i.e. we compute the cross-power spectrum of ${\kappa_{\rm SZ} \times \delta_g}$ relative to that of  ${}\kappa_{\rm CMB} \times \delta_g$:
\begin{equation}
 b_{\ell} = \frac{C_{\ell}^{\kappa_{\rm SZ}\times\delta_g}}{C_{\ell}^{\kappa_{\rm CMB}\times\delta_g}}\,,
  \label{eq:corrbias}
\end{equation}
where $\kappa_{\rm SZ}$ is the map of the residual tSZ-induced lensing convergence field obtained from Eq.~\eqref{eq:kappasz}, while ${\kappa_{\rm CMB}\equiv \kappa_{\rm SMICA-noSZ}}$ is the "tSZ-free" \emph{Planck} CMB lensing convergence map that was derived from the tSZ-deprojected \emph{Planck} \textsc{smica-nosz} CMB map, and $\delta_g$ is the SDSS galaxy density contrast map described in Sect.~\ref{bossdata}.

The relative bias due to spurious correlations between the \emph{Planck} residual tSZ-induced lensing field and the SDSS galaxy density field is plotted in Fig.~\ref{fig:reMH18} as \emph{red-coloured} bandpowers. The cross-power spectrum shows the scale-dependent transitional shape of the tSZ-induced bias as anticipated from Fig.\,\ref{fig:stack}. Our result derived from \emph{Planck} and SDSS data is found to be consistent with theoretical projections in \cite{mh18} and \cite{van_engelen2014} using model simulations, with a negative correlation bias at low multipoles $\ell < 1500$ and a positive correlation bias at high multipoles $\ell > 1500$.

Since the CMB lensing signal is cancelled out through the map difference in Eq.~\eqref{eq:kappasz}, the residual convergence map is dominated by the spurious lensing field $\kappa_{\rm SZ}$ induced by residual tSZ contamination in the tSZ-contaminated \emph{Planck} \textsc{smica} lensing map. In comparison, we also show results in which the  $\kappa_{\rm SZ}$ map is computed from the difference between the \emph{Planck} lensing convergence derived from the cluster-masked \textsc{smica} map and the one derived from the tSZ-deprojected \textsc{smica-nosz}  map, i.e. ${\widetilde{\kappa}_{\rm SZ} = \kappa_{\rm Cluster-masked} -\kappa_{\rm SMICA-noSZ}}$ (see \emph{green-coloured} bandpowers in Fig.~\ref{fig:reMH18}). The spurious tSZ-induced lensing field obtained in this case is thus mitigated by cluster masking.  In all cases, the $\kappa_{\rm CMB}$ in Eq.~\eqref{eq:corrbias} is from the tSZ-deprojected \textsc{smica} lensing map.  The quantity in Eq.~\eqref{eq:corrbias} plotted in Fig.~\ref{fig:reMH18} thus gives the fractional bias in the CMB lensing-galaxy cross-spectrum induced by tSZ projections in each of the \emph{Planck} CMB lensing maps. The data in \emph{black} show the $1\sigma$ sample variance obtained by cross-correlating the two variations of \textsc{smica} lensing maps with 1000 MC LSS maps generated by randomizing SDSS galaxy locations.

In contrast to MC fluctuations, which on average are consistent with zero within the $1\sigma$ sample variance, the spurious tSZ-induced lensing from the \textsc{smica} tSZ-contaminated map (\emph{red}) causes a negative percentage bias of $\sim2.5\%$ on the cross-spectrum at large scales of  $\ell\lesssim500$, crosses the sign at scales of $500\lesssim\ell\lesssim1500$ with a slight negative bias of  $\sim1\%$, and shows a  positive bias of $\sim9\%$ at small scales of $\ell\gtrsim1500$. We quantify the  detection significance of the fractional bias in the cross-spectrum by computing the signal-to-noise ratio (SNR) over each multipole bin. In each bin $i$, the SNR is calculated as the mean signal $\overline{c_i}$ (\emph{red}) over the $1\sigma$ sample variance $\overline{\sigma_i}$ (\emph{black}) of the same sign  within that particular multipole bin so that
\begin{equation}\label{sni}
 \frac{\rm S}{\rm N}\bigg\vert_i  = \frac{\overline{c_{i}}}{\overline{\sigma_i}}\,.
\end{equation}
The overall SNR  is the square root of the quadratic sum of all multipole bins such that
\begin{equation}\label{sntot}
 \frac{\rm S}{\rm N}\bigg\vert_{\rm tot}  = \sqrt{\sum_i^5\left(\frac{\rm S}{\rm N}\bigg\vert_i \right)^2}
\end{equation}
The SNR for the different multipole ranges are listed in the middle column of Table\,\ref{tab:MH18SN}. The negative bias at large scales of $100<\ell<500$  is detected with a $7.4\sigma$ significance, and the positive bias at $1500<\ell<2500$ is detected with a $12\sigma$ significance. Over all multiple scales, we detect the tSZ-induced lensing bias with a $14.6\sigma$ significance. 

In contrast with the tSZ-contaminated \textsc{smica} lensing map, the cross-spectrum bias (\emph{green} in Fig.\,\ref{fig:reMH18}) caused by the use of the cluster-masked \textsc{smica} lensing map is reduced thanks to the masking of clusters where most of the tSZ emission comes from. Nevertheless, the cross-spectrum still gives a negative bias of $1.8\%$ at large scales of $\ell\lesssim500$ and transits to a positive bias of $3.1\%$ at   $1500\lesssim\ell\lesssim2500$.  The detection significance of the bias in this case is quantified in the last column of Table\,\ref{tab:MH18SN}.  Compared with the MC fluctuation, the cross-spectrum still detects a $5.4\sigma$ significance at $100<\ell<500$ and  a $4.1\sigma$ significance at $1500<\ell<2500$, yielding a total of $7.4\sigma$ detection. 
These results demonstrate that masking galaxy clusters in the CMB map prior to CMB lensing field estimation, a common approach in the literature \citep[e.g., ][]{boc+19},  is clearly not sufficient to eliminate the tSZ-induced lensing bias on the cross-correlation between CMB lensing and LSS data. Diffuse tSZ emission and undetected clusters seemingly cause a non-negligible lensing bias on CMB lensing-galaxy cross-correlations.


\begin{figure*}
 \includegraphics[width=0.48\hsize]{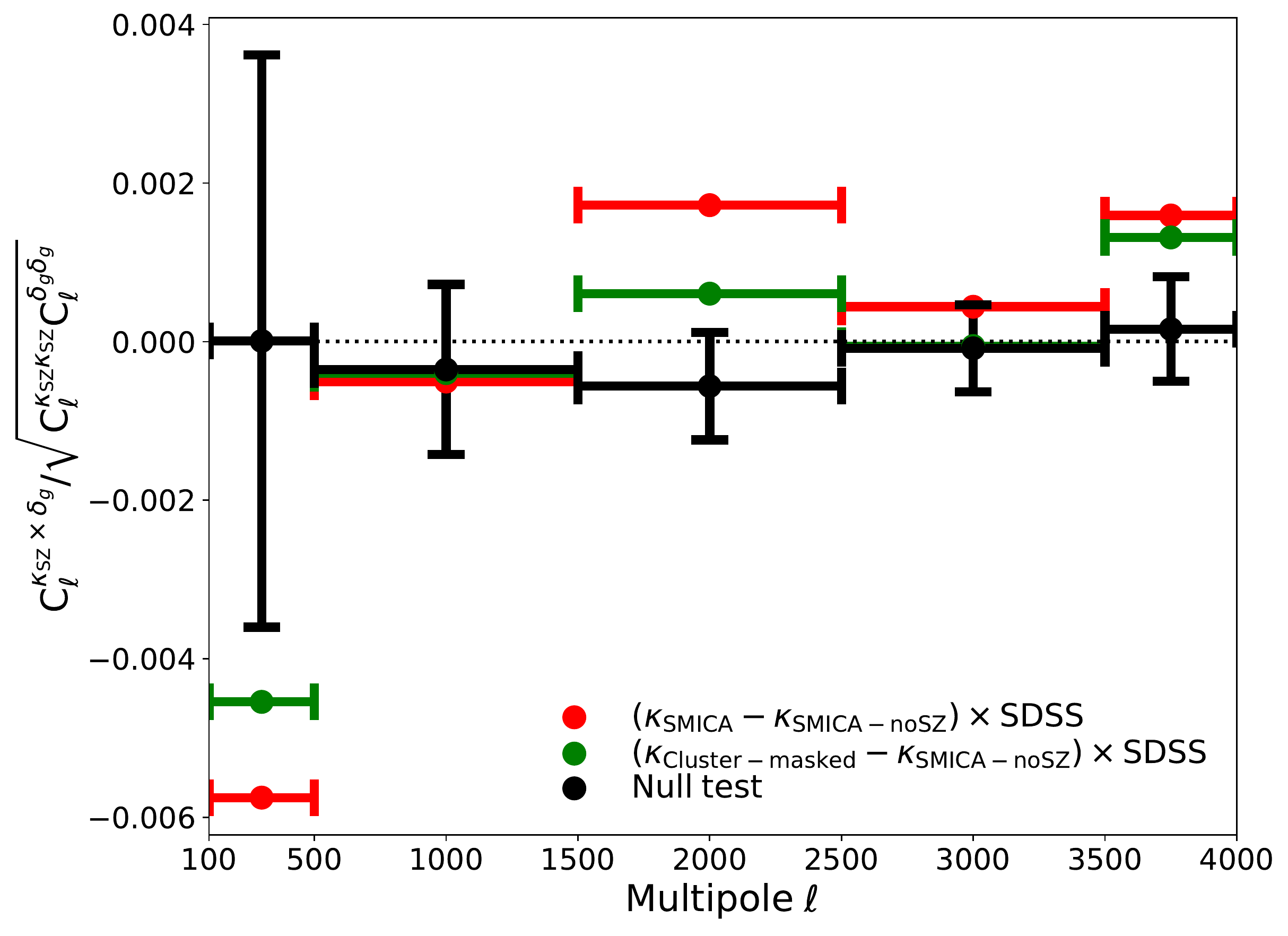}
 \includegraphics[width=0.48\hsize]{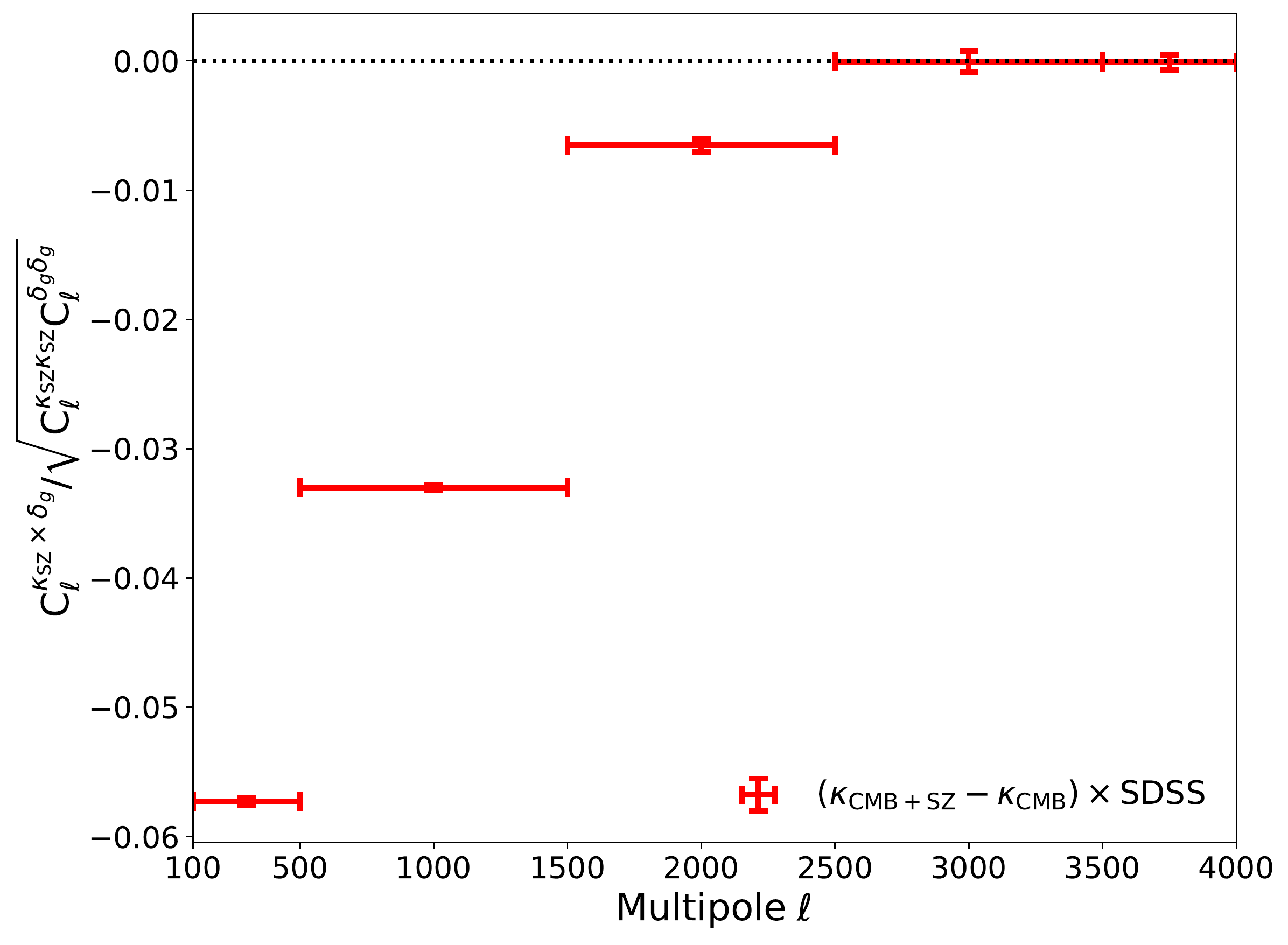}
 \caption{\emph{Left:} Cross-correlation coefficient across multipoles between the SDSS MphG map and the \emph{Planck} residual tSZ-induced lensing convergence map $\kappa_{\rm SZ}$.   The $\kappa_{\rm SZ}$ map in the \emph{red} data set is from the difference between the tSZ-contaminated \textsc{SMICA} and tSZ-deprojected \textsc{SMICA} lensing  maps ($\kappa_{\rm SMICA}-\kappa_{\rm SMICA-noSZ}$). The \emph{green} data set is from the difference between the cluster-masked  \textsc{SMICA} and SZ-deprojected \textsc{SMICA} lensing  maps ($\kappa_{\rm Cluster-masked}-\kappa_{\rm SMICA-noSZ}$).  The horizontal bars indicate the multipole bin width. The black error bars give the average and $1\sigma$ sample variance computed from 1000 MC simulations with randomised SDSS galaxy locations.  \emph{Right}: Cross-correlation coefficients of the SDSS MphG map with 1000 MC lensing convergence difference maps ($\kappa_{\rm CMB+SZ}-\kappa_{\rm CMB}$) derived from pure lensed CMB realisations ($\kappa_{\rm CMB}$) and artificial 100\% tSZ-contaminated lensed CMB realisations ($\kappa_{\rm CMB+SZ}$). The data points indicate the mean of MC cross-correlation coefficients and the $1\,\sigma$ uncertainty from MC sample variance. }
\label{fig:diffxsdss}
\end{figure*}

\begin{table}
\centering
\begin{tabular}{lcc}
\hline
\hline
\multirow{2}{*}{Multipole Range} & \multicolumn{2}{c}{S/N} \\
& tSZ-contaminated & Cluster-masked\\
\hline
{[100, 500]} & 1.6 & 1.2\\
{[500, 1500]} & 0.4 & 0.3\\
{[1500, 2500]} & 14.6 & 5.1\\
{[2500, 3500]} & 1.0 & 0.1\\
{[3500, 4000]} & 1.9 & 1.6\\
\hline
[100, 4000] & 14.8 & 5.5\\
\hline
\hline
\end{tabular}
\caption{Detection significance of the excess correlation between the spurious tSZ lensing field in the \emph{Planck} CMB lensing map and the SDSS galaxies over different ranges of multipoles. The S/N is based on the left panel of Fig.~\ref{fig:diffxsdss}. The middle column shows the results from the tSZ-contaminated lensing map, while the last column shows those from the cluster-masked lensing map. }
\label{tab:SNtab}
\end{table}

\subsection{Cross-correlation between \emph{Planck} residual SZ-induced lensing  and SDSS galaxies}
\label{sec-diffxsdss}

In order to quantify the amount of spurious lensing due to residual tSZ emission in the \emph{Planck} CMB lensing products, we compute the dimensionless Pearson cross-correlation coefficient between the SDSS galaxy density field and the residual tSZ-induced lensing convergence field across multipoles as 
\begin{equation}
  \rho_{\ell} = \frac{C_{\ell}^{\kappa_{\rm SZ}\times \delta_g}}{\sqrt{C_{\ell}^{\kappa_{\rm SZ}\kappa_{\rm SZ}}C_{\ell}^{ \delta_g \delta_g}}}\,.
  \label{eq:diffxsdss}
\end{equation}
Again, we computed two sets of residual tSZ-induced lensing field, $\kappa_{\rm SZ}$, from i) the difference between the \textsc{smica} tSZ-contaminated  and the \textsc{smica} tSZ-deprojected lensing maps; and ii)  the difference between the \textsc{smica} cluster-masked and the  \textsc{smica}  tSZ-deprojected  lensing maps. 

The left panel of Fig.~\ref{fig:diffxsdss} shows the scale-dependent correlation between SDSS galaxies and the spurious tSZ-induced lensing field arising from either the \textsc{smica} tSZ-contaminated CMB lensing map (\emph{red}) or the cluster-masked CMB lensing map (\emph{green}). The $1\sigma$ sample variance (\emph{black}) is computed by cross-correlating the $\kappa_{\rm SZ}$ map with 1000 MC maps of randomly located SDSS galaxies. For both the tSZ-contaminated and cluster-masked \emph{Planck} lensing products, an excess anti-correlation with SDSS galaxies is observed at $\ell\lesssim500$ and an excess positive correlation is observed at $\ell\gtrsim1500$, with the transition of sign happening between $\ell = 500$ and $\ell = 1500$. The spurious lensing field due to residual tSZ effect from the cluster-masked \textsc{smica} map (\emph{green}) show relatively less (anti-)correlation with SDSS galaxies compared to the tSZ-contaminated  \textsc{smica} map (\emph{red}), as expected.  Nevertheless, masking tSZ clusters in CMB maps does not get rid of the spurious tSZ-induced lensing field, as a $>1\sigma$ (anti-)cross-correlation signal is  still detected with the same typical multipole-dependent shape. This result again emphasises the importance of using proper tSZ-deprojected CMB lensing maps in place of cluster masking for cross-correlation analysis with LSS data.  The scale-dependent transitional correlation shown in Fig.~\ref{fig:diffxsdss} consolidates our results from map stacking in Fig.~\ref{fig:stack} and cross-power spectrum bias in Fig.\,\ref{fig:reMH18}. 

Following the same procedure as in section\,\ref{sec-kxsdss}, we quantify the significance of the detection of the spurious cross-correlation signal between SDSS galaxies and the residual tSZ-induced lensing field in Fig.~\ref{fig:diffxsdss} using Eq.~\eqref{sni}-\eqref{sntot}. The results for different multipole bins are listed in Table\,\ref{tab:SNtab}. The middle column shows results for the tSZ-contaminated \emph{Planck} lensing map, corresponding to the \emph{red} curve in the left panel of Fig.\,\ref{fig:diffxsdss}. The anti-correlation at large scales ($\ell<500$) is detected with $1.6\sigma$ significance, while the excess positive correlation at small scales ($1500<\ell<2500$) is detected with highest significance at $14.6\sigma$, thus giving an overall $14.8\sigma$ detection significance over the full range of multipoles from $\ell = 100$ to $\ell = 4000$. In comparison, the last column shows results for the cluster-masked \emph{Planck} CMB lensing map. The SNR of the tSZ-induced cross-correlation in this case,  although less significant, is yet giving a 5.5$\sigma$ detection in total. These results again demonstrate that while masking galaxy clusters for CMB lensing maps helps mitigating the excess correlation with LSS data, it is still insufficient to get rid of such spurious correlations.

\begin{figure*}
 \includegraphics[width=0.48\hsize]{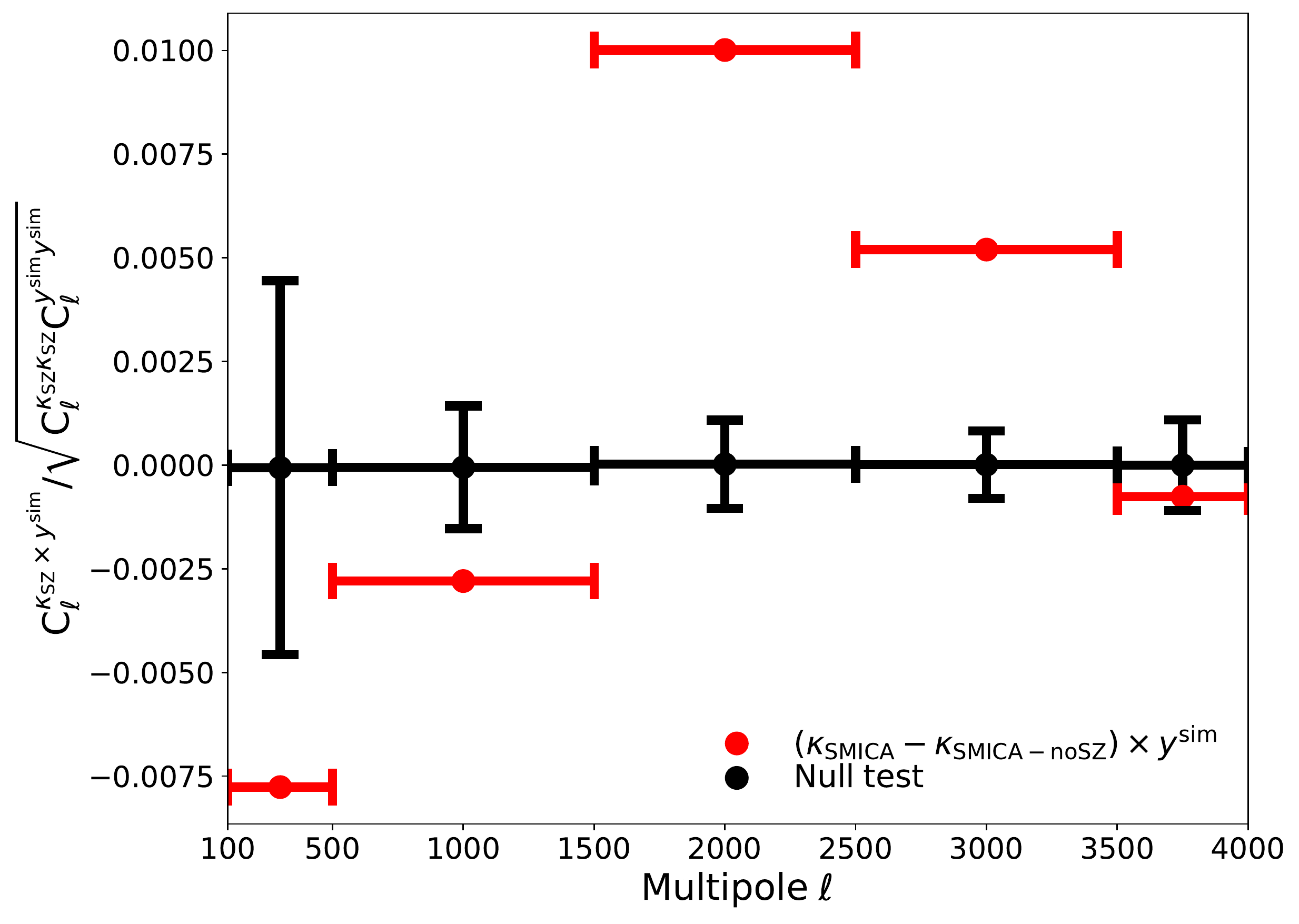}
 \includegraphics[width=0.48\hsize]{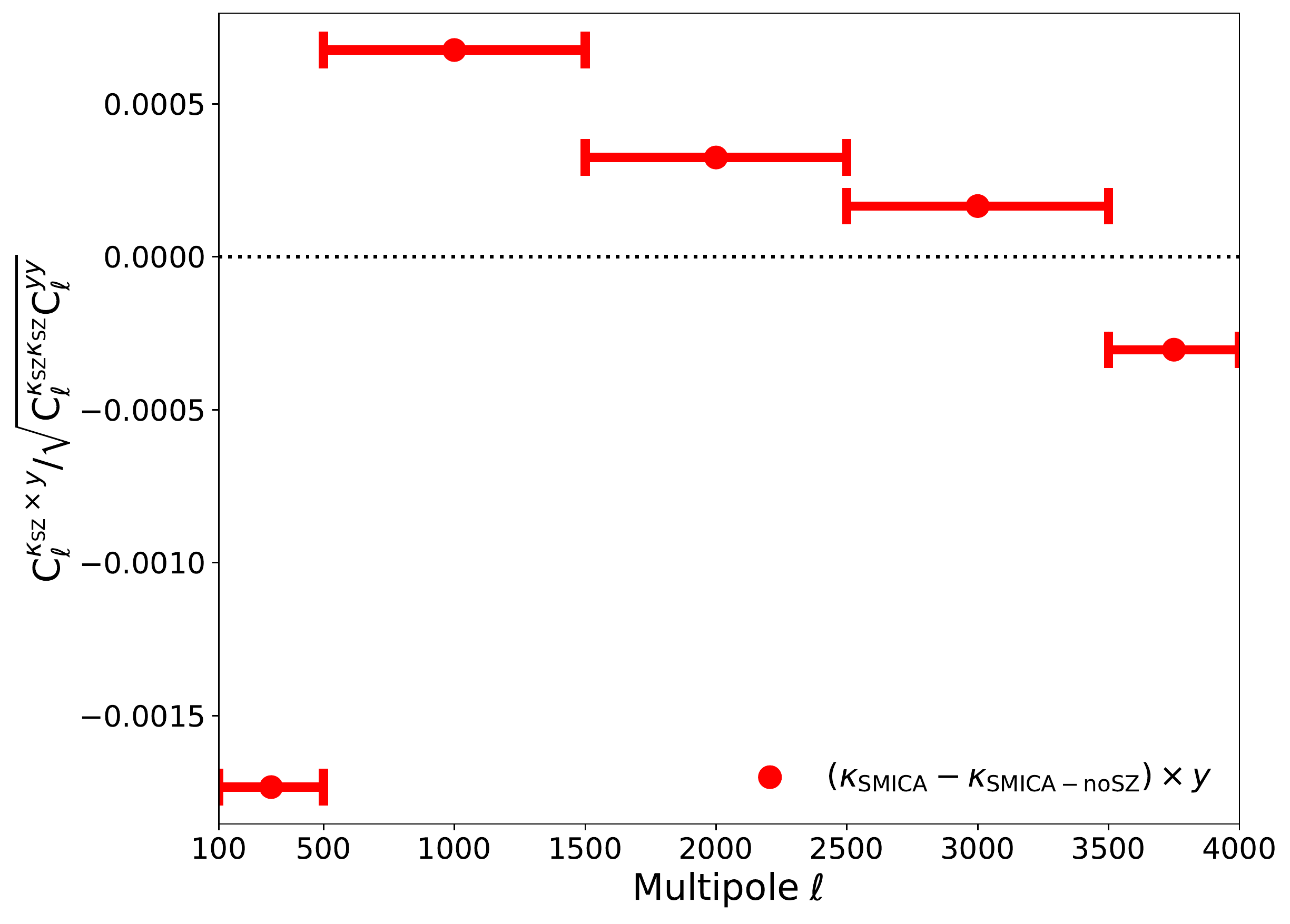}
 \caption{\emph{Left:} Cross-correlation coefficient across multipoles between the \emph{Planck} SZ catalogue $y$-map  and the \emph{Planck} residual tSZ-induced lensing convergence map ($\kappa_{\rm SZ} = \kappa_{\rm SMICA}-\kappa_{\rm SMICA-no SZ}$; \emph{red}). The black data points give the mean and $1\sigma$ sample variance computed from 1000 MC simulations with randomised SZ cluster locations. \emph{Right:}  Cross-correlation coefficient across multipoles between the \emph{Planck} \textsc{nilc} thermal SZ $y$-map and the \emph{Planck} residual tSZ-induced lensing convergence map $\kappa_{\rm SZ}$. Note that the \emph{Planck} \textsc{nilc} $y$-map has an angular resolution of 10\,arcmin, so the \emph{Planck} lensing convergence map has been smoothed accordingly.}
\label{fig:diffxtsz}
\end{figure*}

To further validate the source of the spurious cross-correlation signal from \emph{Planck} lensing data, we cross-correlate the SDSS galaxy density map with 1000 MC simulations of tSZ-induced lensing convergence maps which we computed from the difference between artificial 100\% tSZ-contaminated CMB lensing convergence maps and pure CMB lensing convergence maps, as described in section\,\ref{simsec}. The resulting cross-correlation coefficient across multipoles is shown in the right panel of Fig.~\ref{fig:diffxsdss}. The \emph{red} data points show the mean and $1\sigma$ uncertainty of the MC cross-correlation coefficients. A clear negative excess is detected at  angular scales with $\ell\lesssim2500$, confirming that the observed anti-correlation using the \emph{Planck} lensing data at large angular scales in the left panel of Fig.~\ref{fig:diffxsdss} is indeed caused by the spurious lensing field induced by tSZ residuals which cross-correlates with SDSS galaxies. The amplitude of the negative excess from the MC simulation is more significant than that from the \emph{Planck} lensing data in the left panel. This is because the tSZ contamination in \emph{Planck} data has already been partly suppressed through component separation, while the MC simulation includes the full tSZ emission.   However, the expected transition to positive correlation as shown in the \emph{Planck } data is not clear in the MC simulation. This is potentially due to the incompleteness of the \emph{Planck} SZ catalogue used in the MC simulation, which does not include every galaxy clusters nor the diffuse tSZ emission all across the sky. 


\subsection{Cross-correlation between \emph{Planck} residual SZ-induced lensing and \emph{Planck} SZ maps}
\label{sec-kxsz}

As further evidence, we cross-correlate the spurious tSZ-induced \emph{Planck} lensing convergence map, $\kappa_{\rm SZ}$, with \emph{Planck} tSZ Compton-$y$ maps. We consider both the \emph{Planck} \textsc{nilc} $y$-map and another $y$-map generated from the \emph{Planck} SZ catalogue as described in section\,\ref{szcat}. The cross-correlation coefficient in this case is defined as
\begin{equation}
  c_{\ell} = \frac{C_{\ell}^{\rm \kappa_{\rm SZ} \times y}}{\sqrt{C_{\ell}^{\rm \kappa_{\rm SZ}\kappa_{\rm SZ}}C_{\ell}^{yy}}}\,,
  \label{eq:diffxsdss}
\end{equation}
and is shown in the left panel of Fig.~\ref{fig:diffxtsz} (\emph{red}) for the catalogue $y$-map and in the right panel of Fig.~\ref{fig:diffxtsz} (\emph{red}) for the \emph{Planck} \textsc{nilc} $y$-map. The $1\sigma$ sample variance (\emph{black}) is  computed by cross-correlating the  $\kappa_{\rm SZ}$ map with 1000 MC SZ catalogue maps of randomised cluster locations. The cross-correlation coefficient  between the residual lensing convergence map  $\kappa_{\rm SZ}$ and the \emph{Planck} SZ catalogue map $y^{\rm sim}$ (\emph{left}) shows anti-correlation at large angular scales and transits to positive correlation  at small angular scales $\ell\gtrsim1000$. This is consistent with the transitional shape of the correlation between the \emph{Planck} tSZ-induced lensing map and the SDSS galaxies in Fig.~\ref{fig:diffxsdss}. 

In the right panel of Fig.~\ref{fig:diffxtsz} we computed the cross-correlation between the tSZ-induced \emph{Planck} lensing convergence map, $\kappa_{\rm SZ}$, and the \emph{Planck}  \textsc{nilc} thermal SZ $y$-map \citep{planck2015sz}, which traces all the hot gas across the sky both from compact galaxy clusters and diffuse regions between clusters.
The tSZ-induced \emph{Planck} lensing convergence map $\kappa_{\rm SZ}$ was smoothed down to 10\,arcmin prior to cross-correlation in order to match the angular resolution of the \emph{Planck} thermal SZ $y$-map.
The result of such cross-correlation is shown in the right panel of Fig.~\ref{fig:diffxtsz}.
The amplitude of the correlation is quite smaller compared to that of the left panel of Fig.~\ref{fig:diffxtsz} where the \emph{Planck} SZ catalogue map was used in place of the \emph{Planck}  \textsc{nilc} thermal SZ $y$-map. This is due to the $10'$ beam smoothing effect in this case, and diffuse foreground and noise contamination in the \emph{Planck} thermal SZ $y$-map.
 Nevertheless, the cross-correlation coefficient between the spurious tSZ-induced lensing map $\kappa_{\rm SZ}$ and the \emph{Planck} thermal SZ $y$-map  recovers the similar transitional correlation than that between  $\kappa_{\rm SZ}$ and the SDSS galaxy density map.


\section{Conclusions}\label{dissec}
In this paper, we conducted a systematic study on the impact of residual tSZ-induced lensing contamination on the cross-correlation between \emph{Planck} CMB lensing maps and LSS tracers, such as SDSS galaxy density and \emph{Planck} tSZ maps. Using the three \emph{Planck} 2018 CMB lensing products in temperature (tSZ-contaminated, tSZ-masked, and tSZ-deprojected), we estimated the residual tSZ-induced lensing convergence field $\kappa_{\rm SZ}$ from either the difference map ${\kappa_{\rm SMICA}-\kappa_{\rm SMICA-no SZ}}$ (for the tSZ-contaminated  \emph{Planck} CMB lensing map) or the difference map ${\kappa_{\rm Cluster-masked}-\kappa_{\rm SMICA-no SZ}}$ (for the cluster-masked \emph{Planck} CMB lensing map).

Through cluster stacking analysis, we highlighted, for the first time on maps, the scale-dependent sign-changing correlation between the spurious tSZ-induced lensing convergence field and the SDSS galaxy density field (Fig.~\ref{fig:stack}). The spurious lensing convergence field induced by residual tSZ contamination in CMB maps shows an increment of convergence in the central part of the clusters and a decrement of convergence in the cluster outskirts.

We also reported a detection of the tSZ-induced lensing bias in the cross-power spectrum between the tSZ-contaminated \emph{Planck} CMB lensing map and the SDSS galaxy density map, causing a $\sim2.5\%$ deficit of power at low multipoles and a $\sim9\%$ excess of power at high multipoles (Fig.~\ref{fig:reMH18} and Table~\ref{tab:MH18SN}). We also showed that the tSZ-induced lensing bias on the CMB lensing-galaxy cross-power spectrum persists to a lower extent even after having masked out confirmed galaxy clusters in the  \emph{Planck} CMB temperature map prior to lensing reconstruction (Fig.~\ref{fig:reMH18}  and Table~\ref{tab:MH18SN}). 

Cross-correlations between \emph{Planck} spurious tSZ-induced lensing convergence $\kappa_{\rm SZ}$ and SDSS galaxies show a transitional behaviour across the multipoles, with significant anti-correlation at large angular scales below $\ell\sim 500$, crossing zero at $500\lesssim\ell\lesssim1500$,  and positive correlation at small angular scales above $\ell\sim1500$ (Fig.~\ref{fig:diffxsdss}).  The spurious excess cross-correlation signal due to residual tSZ-induced lensing is detected at  $14.8\sigma$ significance when integrating over the whole range of multipoles, and still persists to $5.5\sigma$ significance after cluster masking (Table~\ref{tab:SNtab}).

As further evidence, we also cross-correlated the \emph{Planck} residual tSZ-induced lensing map $\kappa_{\rm SZ}$ with tSZ galaxy clusters from both a catalogue map based on the \emph{Planck} SZ catalogue and the \emph{Planck} \textsc{nilc} $y$-map. The same transitional correlation signal is detected as in the case using the SDSS galaxies (Fig.~\ref{fig:diffxtsz}). 

Our results on \emph{Planck} and SDSS data are consistent with theoretical projections from \cite{van_engelen2014}  and numerical simulations  from \cite{mh18}, but also with recent cross-correlation studies using different CMB and LSS data sets \citep[e.g.][]{boc+19}.

Given that masking known galaxy clusters in the CMB temperature map prior to CMB lensing reconstruction still leaves a non-negligible tSZ-induced lensing bias on CMB lensing-LSS cross-correlations, we advocate for the use of CMB lensing maps derived from CMB temperature maps in which the tSZ effect has been fully deprojected for reliable cross-correlation studies. 
Spectral deprojection of tSZ effect through multi-frequency component separation \citep[e.g.][]{Remazeilles2011a} has the merit of eliminating any spurious extragalactic contribution to the CMB lensing field that arises from galaxy clusters, including unknown clusters and diffuse tSZ emission.

The tSZ-induced scale-dependent lensing bias can impact the constraints on cosmological parameters involving cross-correlations between CMB lensing and LSS, such as  the amplitude $\sigma_8$ of the matter power spectrum \citep[e.g.,][]{ogp+19, boc+19,sms+20} and the linear growth of structure, $D(z)$, as a function of redshift  \citep[e.g.,][]{Giannantonio2016, mb19}.
  In particular, \cite{boc+19} showed on simulations that the tSZ contamination of CMB lensing, if not properly taken into account in the cross-correlation analysis, could significantly bias the constraints in the $\Omega_m-\sigma_8$ plane derived from the cross-correlation between SPT lensing data and DES galaxies. In addition, the multipole-dependent bias induced by tSZ on the CMB lensing-galaxy cross-power spectrum can lead to inaccurate measurements of the scale-dependent galaxy bias at different redshifts  \citep[e.g.,][]{gvh+18}.

With upcoming CMB experiments \citep{CMBS4, SO19} and galaxy surveys \citep{LSST2009, Euclid2011} of finer resolution and larger sensitivity, we stress that 
future constraints on cosmological parameters will be even more sensitive to such cross-correlation biases induced by residual tSZ contamination in CMB lensing maps.
As forecasted by \cite{ssf20}, who computed the cross-correlation signal between the expected CMB lensing map from the Simons Observatory and expected LSS data from LSST, the tSZ-induced correlation excess will be larger than the statistical uncertainty in the CMB lensing-galaxy cross-power spectrum. Therefore, one must take extra care about the tSZ cluster residuals for future CMB lensing data, and \emph{constrained} foreground cleaning algorithms \citep{Remazeilles2011a, Abylkairov2021} should be implemented wherever possible to eliminate extragalactic foreground contamination from CMB lensing observables.

\section*{Acknowledgements}
We are grateful to Julien Carron for his crucial  guidance on using the \textsc{LensIt} code and the \emph{Planck} 2018 lensing pipeline. We thank Ant\'on Baleato Lizancos and Anthony Challinor for their generous advice on the CMB lensing simulation and their help on the \textsc{Quicklens} code, which was initially used for our simulation. We also thank Clive Dickinson for his useful comments at the early stages of this work, and Simone Ferraro for constructive remarks on the latest version of the paper. We also thank the anonymous referee for their positive comments and suggestions. MR acknowledges support by the ERC Consolidator Grant CMBSPEC (No. 725456) as part of the European Union’s Horizon 2020 research and innovation program.

\section*{Data Availability}
The \emph{Planck} CMB lensing and SZ data underlying this article are available in the Planck Legacy Archive (\url{https://pla.esac.esa.int}). The SDSS galaxy data underlying this article are available in the SDSS III DR 13 database (\url{https://skyserver.sdss.org/dr13/en/home.aspx}).


\bibliographystyle{mn2e}
\bibliography{journals,lit} 



\bsp	
\label{lastpage}
\end{document}